# Simple and Effective Distributed Computing with a Scheduling Service

Version #1, 21 May 2001
**David M. Mackie, Army Research Laboratory, Adelphi, MD**

**Abstract:** High-throughput computing projects require the solution of large numbers of problems.  In many cases, these problems can be solved on desktop PCs, or can be broken down into independent "PC-solvable" sub-problems.  In such cases, the projects are high-performance computing projects, but only because of the sheer number of the needed calculations.  We briefly describe our efforts to increase the throughput of one such project.  We then explain how to easily set up a distributed computing facility composed of standard networked PCs running Windows 95, 98, 2000, or NT.  The facility requires no special software or hardware, involves little or no re-coding of application software, and operates almost invisibly to the owners of the PCs.  Depending on the number and quality of PCs recruited, performance can rival that of supercomputers.

Many high-performance computing projects are composed of large numbers of computations that, taken individually, are not intrinsically supercomputer-class problems.  However, researchers often use supercomputers for these projects anyway, in order to obtain a high computational throughput.  (Examples of such "high-throughput computing" projects are optimizations, parameter studies, sensitivity analyses, simulated annealing, evolutionary algorithms, ensemble simulations, and combinatorial problems.)  Alternatively, researchers have used distributed computing, tying numbers of workstations together with special software to create a powerful virtual machine.  Taking this idea to its logical conclusion, a recent trend has been to use so-called internet computing, which seeks to make productive use of the spare CPU cycles of millions of home PCs.  The non-profit SETI@home project is perhaps the most famous example, but for-profit companies such as Parabon, Distributed Science, United Devices, and Groove Networks are seeking to cash in.  Each of these options for high-throughput computing has its own advantages, but all share the disadvantage of requiring a sizeable up-front investment of time and money before any results can be obtained.  Small research groups working on a non-sexy problem with a tight budget often fall between the cracks, and wind up grinding out their results for months on the few computers available to them.  In this article, we briefly discuss our approaches to parallelizing a high-throughput project, then show in detail how to quickly set up a simple and effective distributed computing facility using networked Windows PCs.

Our project was to test various methods of optimizing diffractive optical elements (DOEs) so that we could determine the circumstances under which each worked best.  Each of the many optimizations required a large number of finite-difference time-domain (FDTD) calculations.  Although the FDTD computations could be lengthy, they had only moderate memory and storage requirements, which could be handled by any PC not more than a few years old.

The throughput of the investigation could be increased in several ways.  First, one could parallelize the entire investigation, by running different cases on many different computers (or processors, since several of our available computers had multiple CPUs).  This "macroscopic" parallelization approach is guaranteed to be 100% efficient if (but only if) one has the same number of cases as processors, if the cases all have equal computation time, and if the processors are all comparable.  These conditions were not generally applicable for this project, besides which the "macroscopic" approach does nothing to speed up calculations of individual cases.  Despite these limitations, this project made profitable use of this approach on occasion.  There is no doubt that this is the simplest way to perform parallel computing, and its utility is often overlooked.  It can be combined with the distributed computing facility described below.

Second, one could parallelize the FDTD computations, in a "microscopic" parallelization approach.  This had the obvious benefit of speeding up the optimizations of individual cases.  This type of speed-up was one of the desired outcomes of the project, since often one only wants to design a single specific DOE, not a whole slew of them.  The success of this approach varied with the architecture of the computer running the calculations.  Symmetric multi-processing (SMP) architectures (including a 4-processor Dell Poweredge 6300, a 16-processor Origin 2000, and a 64-processor Sun E10000) behaved strangely.  There was an almost perfect speed-up with 2 processors, but allocating additional processors seemed to have no effect.  Auto-parallelizing the same code on a Cray XMP, which has a vector architecture, gave near-linear speed-ups for up to 8 processors (which was as high as we tested).  Probably the FDTD code could have been rewritten to parallelize properly on SMP machines.  However, this experience is an example of one of the pitfalls of "microscopic" parallelization: careful recoding may be required to achieve linear speed-ups.  The other major drawback to this approach became obvious once our group had used up our allotted time on the Cray: it depends upon access to supercomputers.

The only remaining option was to parallelize the optimization of each case, in a "mesoscopic" distributed computing approach.  That is, each FDTD evaluation would be confined to a single CPU, but the hundreds (or thousands) of such evaluations needed for each optimization would be distributed over many computers.  The available computing assets were a hodge-podge of networked PCs running Windows 95, 98, and NT.  It should be understood that the CPUs of many of these machines were actually faster than the CPUs of many supercomputers, so that the potential computing power was substantial.  However, most of these PCs belonged to other people at work, not in our group.  Although they agreed to make their PCs available nights and weekends, they obviously wanted the inconvenience to be minimal.  Lastly, the approach needed to use only standard Windows tools, since we had no skill in Windows systems-level programming.  The remainder of this paper will describe this approach in detail, and will evaluate its success.

There are four parts to our implementation: a batch file to run the optimization executable when a new problem is ready, a few auxiliary files, a few extra lines in the optimization code, and the Windows Task Scheduler to submit the batch file.  The batch file, located on each of the slave computers, is as follows:

DistribOptimize.bat (run by schedule service on each slave computer)

```
subst q: \\MasterPC\CalcDirectory
if exist q:\go.dat DistribOptimize.exe
subst q: /d
exit
```

    The first line maps drive q to a remote directory on the master PC.  The drive letter q is arbitrary; it can be any drive letter not already in use by the slave PC.  In place of MasterPC, one uses the actual network name of the master PC.  (Note that the master PC can simultaneously serve as a slave PC, if desired, since the master PC does very little.)  In place of CalcDirectory, one uses the name of the directory on MasterPC in which calculation results are stored.  In the second line of the batch file, go.dat is an empty file in CalcDirectory that the owner of MasterPC creates when he wishes to run calculations on the slave PCs (as they become available).  If go.dat exists, then the slave PC runs the executable, in this case DistribOptimize.  It is important to note that, in this implementation, the executable file is stored on the slave PCs.  This allows the owners of the slave PCs some assurance that the owner of the master PC is not running a Trojan Horse on their machines.  In a low-trust environment, the owners of the slave PCs could even examine the code and compile it themselves.  In a high-trust environment, the executable file could be stored on the master PC, which would facilitate code changes.  (If the OS of the slave PC is WinNT or Win2000, the batch file can be restricted to reading and writing in a single directory, providing security even with a high-trust implementation.)  When the calculation is completed, or if go.dat does not exist, the batch file frees drive q and exits.

    Minor changes are needed to the optimization code, to take into account that an inhomogeneous group of computers is contributing to the effort, rather than just a single machine.  First, the executable needs a flag to know if the optimization should be initialized, since this should only be done once.  The easiest way to implement this is to initialize separately with the fastest PC available, then start the optimization using many slave PCs.  Second, the executable needs to read in the file with the current best configuration and performance twice – both prior to and after calculating the effect on performance of a configuration change.  Otherwise, the optimization might disregard beneficial configuration changes discovered by other slave PCs.  In theory, determining whether or not the benefits of different changes are mutually exclusive could get rather complicated if they began to pile up on one another.  In practice, for our project, the simple algorithm of (falsely) assuming that beneficial changes were always independent worked well.  Lastly, the executable needs some means to stop.  An easy method is to have the executable stop if go.dat is deleted (either manually, by a batch file on the master PC, or by the executable itself).  If the performance computations (e.g., FDTD) are very lengthy, then they should also check intermittently for this control file, so that a slow slave PC won't unduly delay the master PC from starting the slave PCs on a new case.

    Windows Task Scheduler (WTS) is a standard part of Win95 and Win98.  A secure version is standard with Win2000, and is available on WinNT if Internet Explorer 5.0 or greater is installed.  (To my knowledge, it is not available separately.)  To access WTS, double-click on My Computer, then Scheduled Tasks, then Add Scheduled Task.  This brings up the Scheduled Task Wizard.  Click Next, then Browse.  Find the program you

want WTS to run, which, in our case, is the batch file DistribOptimize.bat that we discussed above. On the next screen, enter the task name (e.g., DistribOptimize), and choose to perform it Daily (as opposed to Weekly or Monthly). On the next screen, choose a start time of 12:00 p.m. and choose to perform it Every Day (as opposed to Weekdays or Every X Days). Leave the Starting Date as is. For the secure version (Win2000 and WinNT), the next screen requests a user name and password. The task will run as if it were started by that user. This feature ensures the safety of the slave PC from Trojan Horses, since the owner of the slave PC can set up a user especially for this project, and restrict that user to only one directory. The next screen has a check box for advanced properties, which you should check, then click Finish. A new screen opens, with three tabs. It should look like Fig. 1. Notice that the Task and Schedule tabs contain all of the information that you just entered during the "wizard" phase, so if you made a mistake, you can correct it. Click on the Schedule tab, then Advanced. (See Figs. 2 and 3.) Check the Repeat Task box, and have the task Repeat Every 10 minutes Until a Duration of 23 hours 50 minutes. (Note that our start time and duration are allowing for the possibility that slave PCs might be available round the clock.) WTS will only start one instance of each scheduled task, so there is no need to worry that these settings will cause problems. Make sure that the box to stop the task is unchecked. Click OK, then the Settings tab. (See Fig. 4.) Uncheck both boxes in the Scheduled Task Completed section, and check both boxes in the Idle Time section. Set the times so that the task will only start if the computer has been idle for at least 60 minutes, and will retry for up to 5 minutes. (The idle time is obviously arbitrary, and the retry time doesn't matter since the task retries every 10 minutes anyway.) Checking the box that stops the task if the computer ceases to be idle insures that the owner of the slave PC is inconvenienced as little as possible. Click OK, and the task has been added.

If you right-click on a previously-saved task in the Task Scheduler window, and choose Properties, the three-tabbed screen reappears. Note that you can disable a task via the Enabled checkbox at the bottom of the Task tab. This is useful if a slave computer's owner wishes to temporarily "opt out" of your distributed computing initiative.

One warning: On WinNT, WTS seems to prefer that the Windows Scheduling Service be already running, rather than on automatic startup, which is the default. To accomplish this, double-click on My Computer, then Control Panel, then Services, then Task Scheduler. Set Startup mode to Manual, and click OK. Issue the command "net start schedule" (without the quotes) from a command prompt. To keep the change user-friendly for the owners of the slave PCs with WinNT, put this command into a batch file in their startup folders.

To recap (see Fig. 5): WTS on each slave PC checks every 10 minutes to see if the PC has been idle for 60 minutes. If it has, WTS starts DistribOptimize.bat, which checks for the existence of the file go.dat on the master PC. If go.dat does not exist, the batch file exits. WTS tries again in 10 minutes. If go.dat does exist, then the executable DistribOptimize.exe runs, comparing its results after each FDTD iteration with the global results on the master PC. The slave PCs make changes to the global results as appropriate. Once the end conditions are fulfilled, go.dat is deleted, signaling DistribOptimize.exe to stop on all slave PCs. The batch file exits and the cycle begins again. If at any time a slave PC ceases to be idle (i.e., mouse movement or key press), then WTS on that PC kills the batch job, which also kills the executable, and control is returned to the owner of that slave PC. The other slave PCs continue with the job, unperturbed.

We found in practice that our "mesoscopic" distributed computing approach worked extremely well, with nearly ideal speedups for up to 10 PCs. (I say "ideal" rather than "linear" because this was an inhomogeneous group of computers.) This was not surprising, in view of Amdahl's Law, because we had purposely minimized the serial component of the problem. There was little competition for memory access, since most of the machines were single-processor. The communication overhead between the master and slave PCs (including read/writes to the master PC's hard drive) was insignificant compared to the lengthy FDTD calculations. If we had been able to scale to very large numbers of PCs, this communication overhead might have become important. The speedup would also have fallen slightly from ideal due to occasional wasted analyses. These could arise from identical analyses (since changes were picked randomly) or analyses to outdated profiles. We could mitigate these problems by tracking changes to configurations, rather than the configurations directly, with a slight increase in the complexity of the optimization routine. The combination of four PII 400 MHz, four PIII 500 MHz, one PIII 1 GHz, and one Athlon 850 MHz CPUs gave us a combined SPECfp95 score of about 160. By way of comparison, an Origin 2000 with sixteen 300 MHz R12k processors scores only 114. Although admittedly our "private supercomputer" was not world-class, it was dedicated to our project 12 hours every workday and continuously on weekends and holidays, all at no charge. We have more participants lined up, so our facility will probably have double the power for future projects.

As a bonus, we found that our "mesoscopic" distributed computing code also worked very well on SMP-type workstations and supercomputers, by treating each processor as a separate computer. After first initializing with one (quick) job, all additional jobs served as slaves. Since each job used only one processor, all were given relatively high priority by the queueing system, and a halt in one job did not require any of the others to wait. In this fashion, we were able to use up our allotted SMP-supercomputer time very rapidly.

Since completing this project, we have discovered that Windows is not the only OS with a scheduling service, so our basic idea could be applied to almost any group of networked computers able to access each other's drives. The signal-file control system could be extended easily to allow multiple master PCs on a network, or different jobs for each slave PC. The best thing about our WTS-based distributed computing system, however, is that it allows a researcher, with minimal fuss, to use his or her bosses' latest powerhouse PCs for something other than email and viewgraphs.

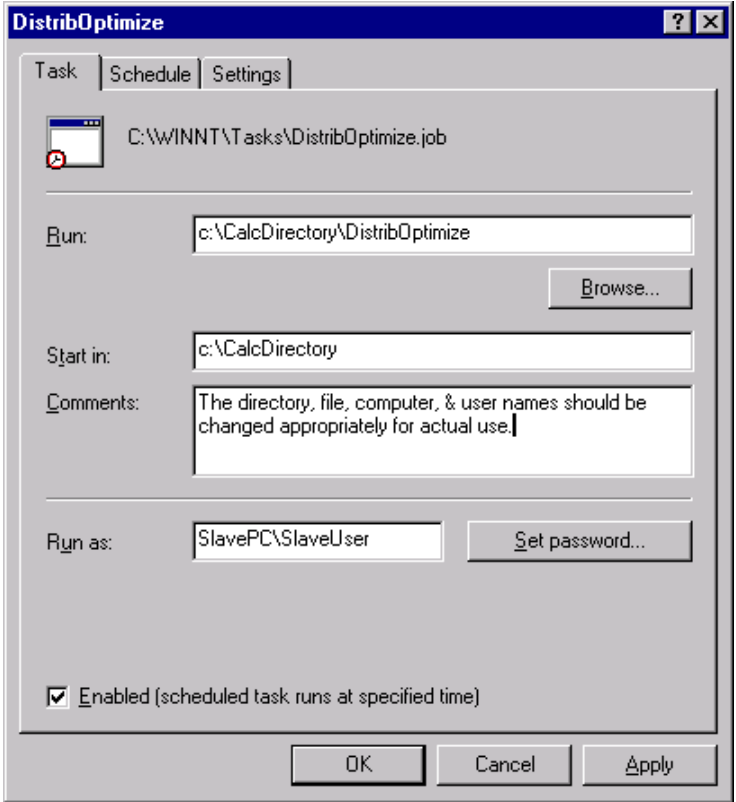

Figure 1. Task Tab of WTS.

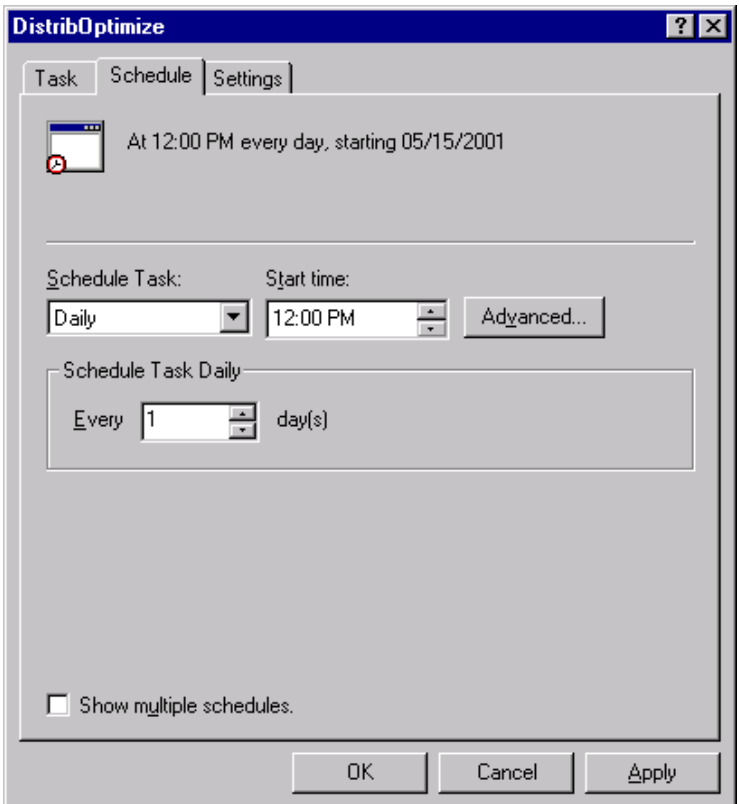

Figure 2. Schedule Tab of WTS.

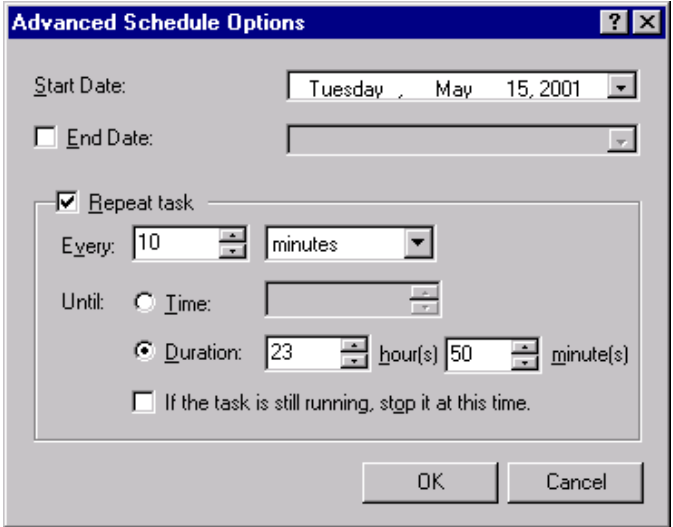

Figure 3. Advanced Schedule Options of WTS.

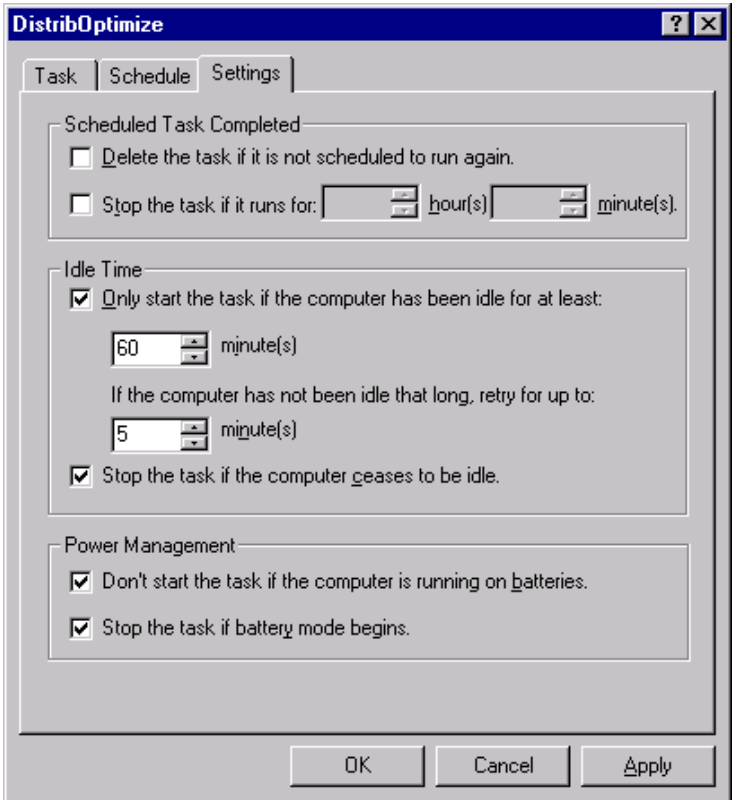

Figure 4. Settings Tab of WTS.

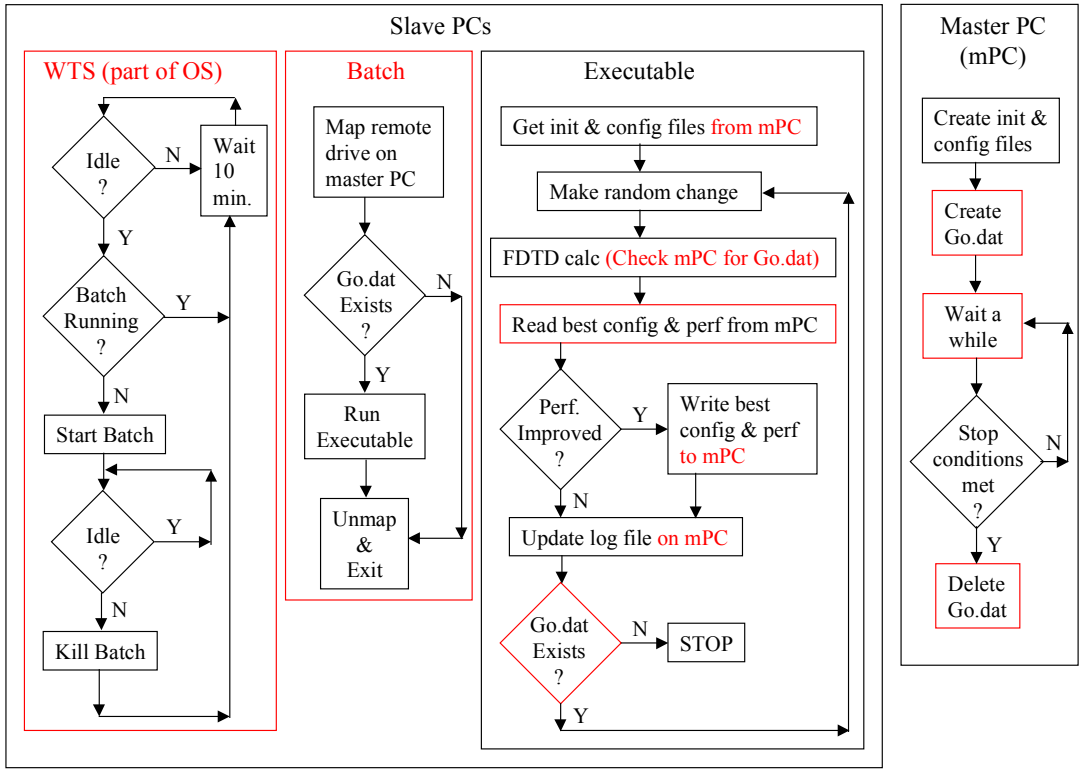

Figure 5. Flow chart for our distributed computing system. Elements unique to the distributed computing version of the program are in a red box or in red text.